\documentclass[12pt,a4paper]{article}

\usepackage{amsmath}
\usepackage{amsfonts}
\usepackage{amssymb}
\usepackage{amsthm}

\newcommand{\SU}{{\rm SU}}
\newcommand{\dd}{{\mathrm d}}      

\providecommand{\abs}[1]{\left\lvert#1\right\rvert} 
 
\newtheorem*{theorem}{Theorem}
\newtheorem*{corollary}{Corollary}

\begin{document}

\title{The Ponzano-Regge model and Reidemeister torsion}

\author{John W. Barrett\\
Ileana Naish-Guzman\\
\\
School of Mathematical Sciences, University of Nottingham\\ 
University Park, Nottingham, NG7 2RD, UK}

\date{}

\maketitle
 
\section{Introduction}
The Ponzano-Regge model of quantum gravity\cite{PR} on a triangulated 3-dimensional manifold was originally presented in terms of a state-sum over representations of SU(2). It is well-known that the analogous model for a finite group can be reformulated in terms of a sum over group elements located on triangles (or dual edges). It is commonly assumed that this is still possible with SU(2). We note that there are subtle questions both about the convergence of the state-sum and also about the fermionic character of the SU(2) representations. To avoid these questions, we present the definition of the Ponzano-Regge model in terms of integrals over elements of SU(2) assigned to the triangles of the triangulation.

There are several different candidates for observables in this model. We define observables specified by giving a conjugacy class in SU(2) to each edge of a graph in the manifold. 
In general there is still a question about whether the resulting integral for the partition function, or `expectation value', of an observable is well-defined. Our first result provides an answer to this question: the criterion for the formula to make sense is that the second twisted cohomology group should vanish at each point of the integration. Our second result says that if this criterion is satisfied, then the resulting expression can be written in terms of the Reidemeister torsion. This proves the independence of the partition function on both the regularisation used in its definition and the triangulation of the manifold. We discuss the particular features of both planar graphs and knots. For a good treatment of the cohomology theory involved, the reader is referred to Dubois\cite{D}.

\section{Definition of the partition function}\label{group}
Let $M$ be a closed 3-manifold with a specified triangulation. The triangulation will have a finite number of simplexes. To specify an observable, we need a graph embedded in $M$, and some data on each edge of the graph. More precisely, let $\Gamma$ be a connected subcomplex consisting of edges and vertices of $M$. For each edge $e$ of this graph, choose a conjugacy class $\theta_e$ of the group $\SU(2)$. The conjugacy class is specified by an angle $\theta_e\in [0,2\pi]$, the angle of the corresponding rotation in Euclidean space. The idea of the Ponzano-Regge model is that it calculates a number $Z$ which is the `expectation value' of this observable. The number $Z$ is often called the partition function, due to the analogy with statistical mechanics.

Let $\Delta_1$ ($\Gamma_1$) be the set of edges of the triangulation (graph).
To define $Z$, it is necessary to pick a regularising subset of edges $T\subset\Delta_1\setminus\Gamma_1$ satisfying the following conditions:
\begin{itemize}
\item
Each connected component of the graph formed by $T$ is a tree (i.e. contains no loops) and is attached to $\Gamma$ at exactly one vertex
\item $T$ is maximal, i.e. visits each vertex of $M$ not contained in $\Gamma$.
\end{itemize}
 
The definition of the partition function is as follows. We use the dual cell decomposition of $M$ in which there is one dual $k$-cell for each $3-k$--cell of $M$. On each dual edge $f$ of $M$, with an arbitrary choice of orientation, there is a variable $g_f\in\SU(2)$ (and $g^{-1}_f$ is assigned to the opposite orientation). This set of variables is called a connection, and given a path consisting of a sequence of oriented dual edges $\gamma=(f_1, f_2, \ldots, f_N)$, there is a holonomy element
\begin{equation*}
H(\gamma) \; = \; g_{f_1}^{\epsilon_{f_1}} \, g_{f_2}^{\epsilon_{f_2}}\, \ldots \, g_{f_N}^{\epsilon_{f_N}}
\end{equation*}
where $\epsilon_{f_i} = \pm1$ according as $f_i$ is traversed in a positive/negative sense (with respect to its orientation).

On each oriented dual face $e$, there is then the  holonomy $h_e=H(\gamma)$ given by the sequence $\gamma$ of dual edges around its boundary. This is well-defined up to conjugation. Finally, the definition uses some delta-functions on $\SU(2)$. The first of these is the delta-function at the identity element $i$, defined by
$$\int_{\SU(2)}\delta(g)F(g) \, \dd g=F(i),$$
for any function $F$, where $\smallint \dd g = 1$.
The second is the delta-function at a conjugacy class $\phi$, given by an ordinary delta-function $\delta(\phi-\theta(g))$. Here, $\theta(g)$ denotes the conjugacy class of $g\in\SU(2)$. 
  
The partition function is obtained by integrating over these variables. 
\begin{equation}\label{partition}
Z(M, \Gamma_\theta) = 
\int\prod_{f\in\Delta_2} \dd g_{f}\prod_{e\in\Gamma_1}\delta(\theta(h_e)-\theta_e)\prod_{e\in\Delta_1\setminus(\Gamma_1\cup T)}\delta(h_e)
\end{equation}

Similar definitions appears in previous works\cite{FLi,FLo2}.

The roles of the various factors in (\ref{partition}) are as follows. The delta-functions for the edges on $\Gamma$ force the holonomy of the connection around that edge of the graph to lie in the conjugacy class $\theta_e$; the delta-functions at the identity force the $g$ variables to give a flat $\SU(2)$ connection on the complement of $\Gamma$; the set of edges $T$ eliminates excess delta-functions, which would otherwise reduce to integrating $\delta^2$ in one of the variables. 

\section{Existence criterion}

\begin{theorem} \label{existence} The partition function (\ref{partition}) exists for a region $\cal R$ of the space of parameters $\{(\theta_1,\theta_2,\ldots)\}$ as a distribution if and only if the second twisted cohomology group $H^2(L,\rho)$ of the graph exterior $L$ is trivial for each flat connection $\rho$ whose conjugacy classes  $(\theta_1,\theta_2,\ldots)$ lie in $\cal R$. 
\end{theorem}
The proof of theorem \ref{existence} and further results below will be given in our forthcoming paper.
In the special case of a planar graph, the existence criterion is always satisfied and so its partition function is always well-defined. It is interesting to consider, in light of our result,  the formula for the tetrahedron graph calculated by Freidel and Louapre \cite{FLo}. For certain values of the parameters, it yields an infinite answer, calling into question the well-definition of this observable. Theorem \ref{existence} tells us that the correct interpretation of the result is as a distribution.
It is the distributional nature of graphs in general that requires the statement of theorem \ref{existence} in terms of a region of parameters.

\section{Invariance of the partition function}

\begin{theorem} If the existence criterion is satisfied then the partition function (\ref{partition}) can be expressed as an integral over the space of flat connections on the graph exterior $L$ with measure given by the Reidemeister torsion, $\mathrm{tor}(L)$.
\end{theorem}

The Reidemeister torsion is known to be a homeomorphism and simple homotopy invariant, and so we have the following

\begin{corollary} The partition function (\ref{partition}) is independent of the choices of triangulation and regularising set $T$.
\end{corollary}

If the graph is a knot $K$, then the partition function vanishes unless all conjugacy classes are equal, so we may, without loss of generailty, take the knot to have a single edge (and a single vertex). If $\theta$ is the associated conjugacy class, then the existence criterion is satisfied for $\theta$ less than a critical angle, $\theta_c(K)$, depending on the knot $K$. For $\theta$ in this range, the partition function is simply a constant times the Reidemeister torsion $\mathrm{tor}(L)$. The simple homotopy invariance of the Reidemeister torsion means we may calculate $\mathrm{tor}(L)$ using the cell complex for $L$ coming from the Wirtinger presentation of $\Pi_1(L)$. Doing so, one obtains
\begin{equation*}
Z(\rm{S}^3, K_\theta) = \rm{const.} \frac{sin^2\theta}{\abs{A_K(e^{i\theta})}^2} \quad\quad 0< \theta < \theta_c(K)
\end{equation*}
where $A_K$ is the Alexander polynomial of $K$. This generalises Barrett's result for the trefoil knot \cite{B}.

\vfill

\end{document}